\documentclass[11pt]{llncs}
\usepackage[a4paper,twoside=false,top=2.32in,bottom=2.69in,left=2.25in,right=2in]{geometry}
\usepackage{times}
\usepackage{graphicx}
\usepackage{mathtools}
\usepackage{amsmath}
\usepackage{tabularx}

\newcommand{\psfigx}[6]{
	\begin{figure}[#5]
	\begin{center}
	\includegraphics[width= #6 in]{#2}
	\end{center}
	\caption{#4}
	\label{fig:#1}
	\end{figure}
}

\begin{document}

\title{An Event~Detection~Approach Based~On Twitter~Hashtags}

\author{
	Shih-Feng Yang and Julia Taylor Rayz}
\institute{
	Purdue University\\
	Computer and Information Technology Department \\
	West Lafayette, IN, 47907\\
           yang798@purdue.edu, jtaylor1@purdue.edu}

\maketitle

\begin{abstract}
Twitter is one of the most popular microblogging services in the world. The great amount of information within Twitter makes it an important information channel for people to learn and share news. Twitter hashtag is an popular feature that can be viewed as human-labeled information which people use to identify the topic of a tweet. Many researchers have proposed event-detection approaches that can monitor Twitter data and determine whether special events, such as accidents, extreme weather, earthquakes, or crimes take place. Although many approaches use hashtags as one of their features, few of them explicitly focus on the effectiveness of using hashtags on event detection. In this study, we proposed an event detection approach that utilizes hashtags in tweets. We adopted the feature extraction used in STREAMCUBE \cite{feng2015streamcube} and applied a clustering K-means approach \cite{lloyd1982least} to it. The experiments demonstrated that the K-means approach performed better than STREAMCUBE in the clustering results. A discussion on optimal K values for the K-means approach is also provided.

\end{abstract}

\section{Introduction} 

Twitter is one of the most popular microblogging services in the world. There are more than 500 million Twitter posts (i.e., tweets) generated per day and around 200 billion per year. The great amount of information within Twitter makes it an important information channel for people to learn and share news. Twitter has several characteristics that distinguish it from news web sites and other information channels \cite{li2012tedas}. First, tweets are created in real-time. For example, a tweet related to a tornado might be written one minute after a user witnessed a tornado was formed. The information could be spread even faster than TV broadcasts. Second, tweets contain information perceived and shared by 'regular' people. When people see gunfire, an earthquake or other events, every witness can share his or her observations and pictures immediately. The information could help to evaluate the actual situation of the events. Third, tweets contain geolocation information. By monitoring tweets about crime events in a specific location, some crimes could be detected immediately.

Hashtag is a popular feature when people use in Twitter. A hashtag is a word or phrase proceeded by “\#”, and is used to identify messages on a specific topic \cite{feng2015streamcube}. For example, “\#ParisAttacks” can be used to indicate the terrorist attacks which happened on the evening of November 13, 2015. It is an important feature that allows researchers to identify the topic of a tweet.

Many researchers proposed event-detection approaches that monitored Twitter data and determined whether special events, such as accidents, extreme weather, earthquakes, or crimes, were happening by analyzing the data on social networks. Data mining techniques related to clustering, classification, and text mining techniques were wildly used in this topic. Although many of them considered hashtag as one of their features, few of them explicitly focused on the effectiveness of hashtag on event detection. In this paper, we proposed an event detection approach that utilized hashtags in tweets. We adopted the feature extraction used in STREAMCUBE \cite{feng2015streamcube} and modified the approach by using the K-means \cite{lloyd1982least} clustering method. Based on our results, we suggest possible improvements for the current research of event detection using Twitter.
\section{Related Works}

Much research on event detection utilized Twitter data to determine whether special events, such as holidays, sport games, earthquakes or crimes, were happening. \cite{middleton2014real} describe a real-time crisis-mapping platform, implemented as an offline service including data-extraction tools for extracting geospatial data. The system contained a parallel geospatial clustering service to continuously cluster spatial areas of high activity. \cite{walther2013geo} proposed an algorithm for geo-spatial event detection on social media streams, which extracted textual features and other attributes from the event candidates, and used a classification component to make a binary decision of whether the candidate was an event. \cite{hua2013sted} presented a semi-supervised system, STED, that can detect target types of events for users in Twitter by extracting action words and named entities from news articles as candidate query words and labeled tweets that contained those words. \cite{ritter2012open} proposed an unsupervised open-domain event-extraction and categorization system, which was a scalable and open-domain approach to extracting and categorizing events from status messages. The approach discovered important event categories and to classify extracted events based on latent variable models. \cite{li2012tedas} proposed a system, TEDAS, to detect new events, to analyze the spatial and temporal patterns of an event, and to identify the importance of events by developing a set of efficient crawler, classifiers, rankers and a prediction module based on crime and disaster related events (CDE) to predict the event locations from the Twitter data. \cite{mathioudakis2010twittermonitor} presented TwitterMonitor that performed trend detection over the Twitter stream by detecting burst keywords by processing tweets with their one-pass real-time algorithm based on queuing theory.  \cite{sakaki2010earthquake} investigated the real-time interaction of events, such as earthquakes, on Twitter, proposed an algorithm to monitor tweets, and detected a target event. The main idea of this approach was taking every Twitter user as a sensor and transforming the problem into an event-based problem on sensory observations. \cite{cataldi2010emerging} proposed a topic detection technique that retrieved the most recent topics expressed by the community in a real-time manner. They discovered the emerging terms that frequently occurred in the specified time interval but were relatively rare in the past. The researchers considered social relationships in the user network to quantify the importance of each analyzed content, then formalized a keyword-based topic graph that connected the emerging terms with their co-occurrent terms. 

The research of hashtag analysis made use of hashtags in tweets to determine the sentiments, preferences and topics of tweets. \cite{feng2015streamcube} proposed STREAMCUBE, which focused on hierarchical spatio-temporal hashtag clustering techniques and generated hashtag clusters for automatically identifying potential events. In order to scale the large amount of Twitter information in different time frames and different areas, the researchers considered both space and time granularity in its database. STREAMCUBE was extended from the traditional data cube. They designed a single-pass clustering algorithm for event identification as well as an event ranking method to find burst events in real-time. \cite{anusha2015twitter} proposed a system contained a topic modeling module to find the score of interest and a sentiment analysis module to detect the polarity. In topic modeling, they adopted Latent Dirichlet Allocation to infer latent topics to which the tweets they collected had belonged. In the sentiment polarity analysis, they used NLTK corpora as training data and the NLTK analysis to decide the sentiment polarity of the tweets. \cite{wang2015hashtag} proposed a hashtag sense induction system to extract a list of words with high node degree and used them to represent a sense of a community. For each hashtag, the system built a list of words as the induced senses of the hashtag. In their implementation, they took the entries in the Wikipedia disambiguation list as Wikipedia senses. \cite{pervin2015hashtag} performed an analysis on the co-occurrence of hashtags. The researchers designed the hypotheses to determine if the popularity of a hashtag increases when it appears along with one or more other similar hashtags. \cite{cepni2014social} modeled the information propagation on Twitter as a sensor network and adopted the communication theories to solve this problem. They considered an event as a sensor which with signal strength. The event estimation was made when an event signal was strong enough at a specific time. \cite{denton2015user} employed user hashtags to capture the description of image content of Facebook users. They utilized the metadata, such as age, gender, home city and country of Facebook users combined with image features extracted from a convolutional neural network algorithm to predict the possible hashtags for images. \cite{wang2014macro} analyzed the hashtag diffusion by macro, the diffusion by the tweet/hashtag properties, and micro perspectives, characterized by Edelman’s topology of influence theory.

Among the above articles, STREAMCUBE \cite{feng2015streamcube} was the only study that mainly focused on tweet clustering based on hashtags to the best of our knowledge. STREAMCUBE proposed a detailed approach for tweet preprocessing, feature extraction, and a single-pass hashtag clustering algorithm. We were interested in how STREAMCUBE would perform when using our Twitter dataset related to the Paris Attacks. We also compare the original STREAMCUBE to our improved version by adopting K-means to for clustering.
\section{Methodology}

The workflow of our study is as follows: first, we collected tweets through Twitter public API and preprocessed the data into features. Second, we implemented the K-means clustering algorithm as the clustering module as well as the clustering algorithm of STREAMCUBE \cite{feng2015streamcube} as another clustering module. Finally, we performed experiments and discussed the performance of the compared clustering approaches.

\psfigx{architecture}{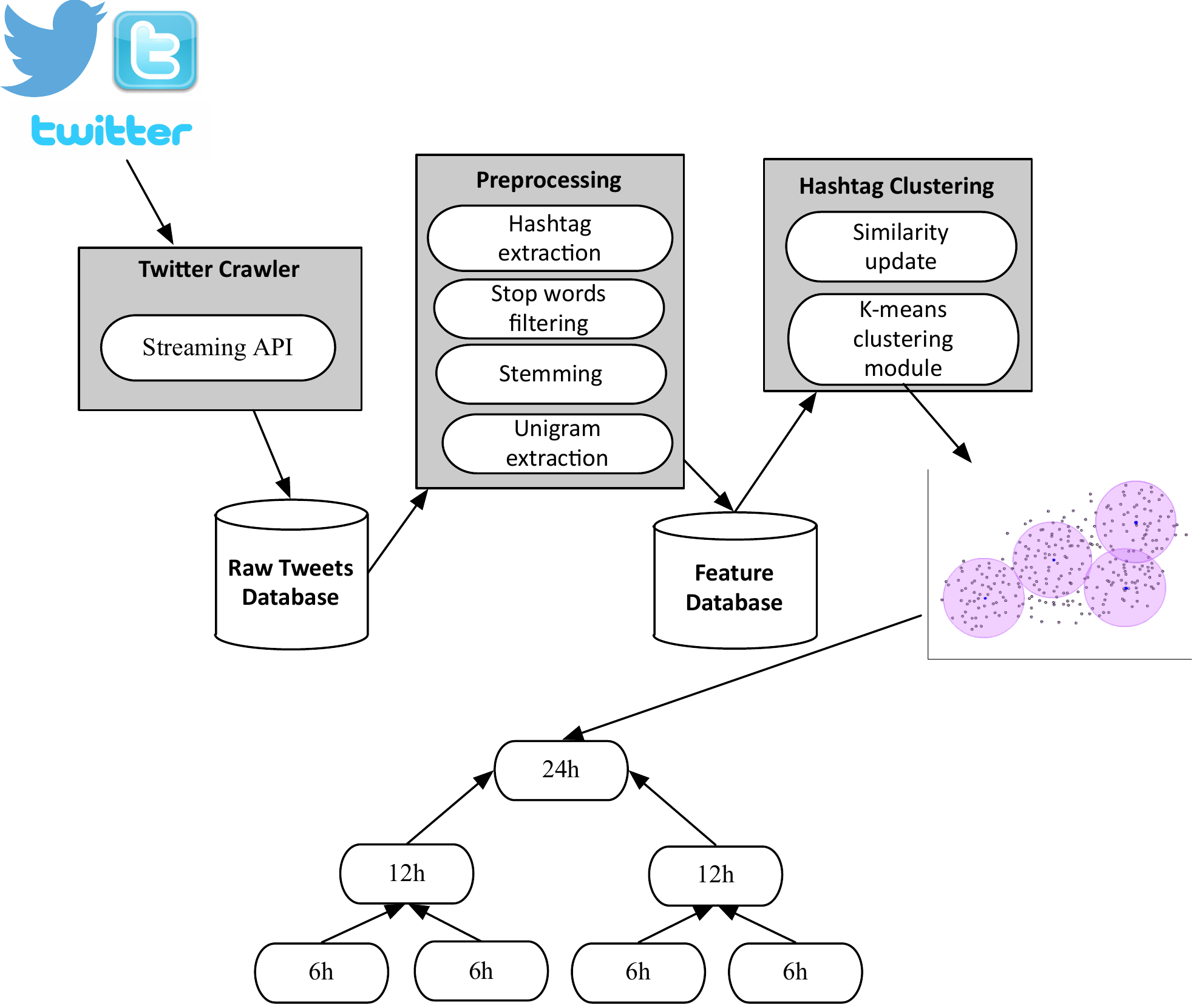}{0}{System architecture}{h}{2}

Figure \ref{fig:architecture} depicts the system architecture of the event detection method based on hashtags. We discuss our data collection and preprocessing for Twitter data in section 3.1. In section 3.2, we described the implementation details of the clustering modules of the K-means approach and the STREAMCUBE approach respectively. The metrics of performance evaluation are described in section 3.3.

\subsection{Data Collection and Preprocessing}

Twitter provided a set of streaming APIs that gave developers low latency access to its global stream of tweets. In this study, we used the Tweepy APIs (https://github.com/tweepy/tweepy), which is a Python library for accessing the Twitter. The APIs enabled us to collect the tweets related to a specific keyword list.

For each tweet, the following properties were collected: created time, number of retweet, text content, mentioned hyperlinks, mentioned hashtags and geographic coordinates. The preprocessing steps in \ref{fig:preprocessing} were used to extract features from the collected tweets: 1) Hashtags in tweets were extracted as unigram features and removed from the original messages. 2) Lowercased all characters in tweets. Removed special characters, stop words, and hyperlinks. 3) All the tweets were stemmed using the Porter stemmer for reducing inflected words to their word stem.

\psfigx{preprocessing}{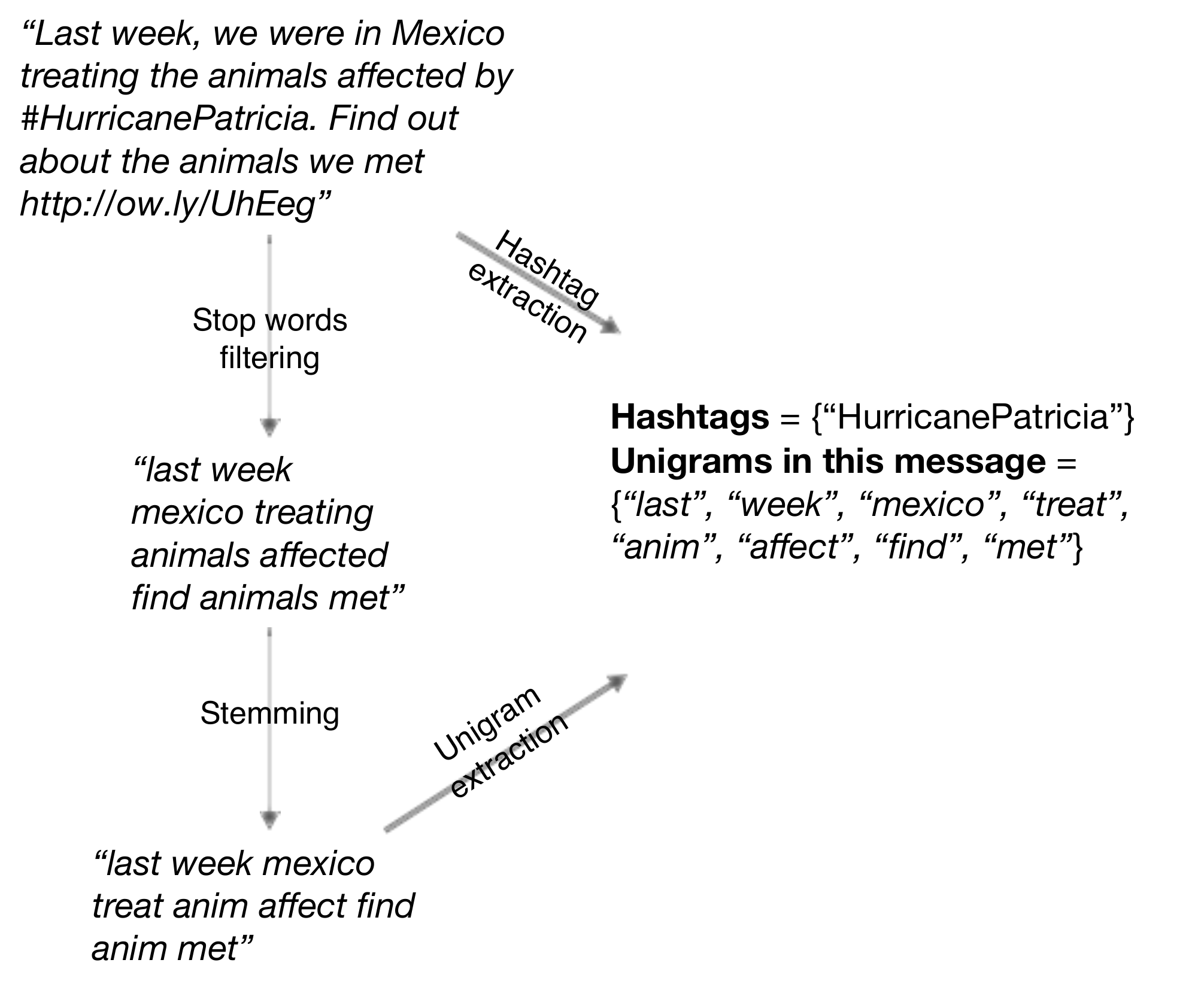}{0}{Tweet preprocessing}{h}{3}

\subsection{Hashtag Clustering}

In the research, we adopted the feature definition of the hashtag clustering approach in STREAMCUBE. In the following sections, we will define the representations of hashtag and introduce how we adopt K-means cluster algorithm.

\subsubsection{Hashtag and Event Representations}

A hashtag $h$ can be considered as a bag of words, which is an aggregation of all the tweets that contain $h$. Using $W$ to denote all the words in our tweets, a hashtag $h$ can be represented as a normalized weighted vector $h_{tweet} = (w_{1}, w_{2}, …, w_{|W|})$ where $w_{i}$ is the weight of the $i$-th word and $\lVert h_{tweet} \lVert = 1$. Additionally, a hashtag h can be considered as a bag of hashtags because many hashtags have co-occurred with other hashtags. Using $H$ to denote the hashtag set, a hashtag $h$ can be represented as a normalized weighted vector $h_{tag} = (h_{1}, h_{2}, …, h_{|H|})$ where $h_{i}$ is the weight of the $i$-th word and $\lVert h_{tag} \lVert = 1$.

By using the above two representation, the distance between two hashtags can be defined. Let $h^{i}_{tweet}$ and $h^{i}_{tag}$ denote the word vector and hashtag vector of the $i$-th hashtag $h^{i}$ . Given two hashtag $h^{1}$  and $h^{2}$, the distance is defined as 

$$sim(h^{1}, h^{2}) = (\alpha^{\frac{1}{2}}h^{1}_{tweet}, \beta^{\frac{1}{2}}h^{1}_{tag})
(\alpha^{\frac{1}{2}}h^{2}_{tweet}, \beta^{\frac{1}{2}}h^{2}_{tag})$$ where $\alpha$ and $\beta$ are two hyperparameters and $\alpha + \beta = 1$. From the above equation, a hashtag can be represented as a vector:

$$h = (\alpha^{\frac{1}{2}}h_{tweet}, \beta^{\frac{1}{2}}h_{tag}) $$

\subsubsection{K-means Clustering}

In the clustering algorithm of STREAMCUBE, the researchers designed a single-pass clustering algorithm aimed at processing data in real time without using an iteration-based algorithm. This issue did not occur in this study because we concentrated on discovering the similarity and dissimilarity between different clustering methods but not the real-time capability.

We adopted the K-means clustering method to find hashtag clusters by using the features described in previous section. We chose K-means because of the following reasons: 1) To identify the clusters, hierarchical clustering and K-means are two well-known cluster algorithms \cite{zhang2012community}. However, considering the great number of features used in our large-scale dataset, K-means is relatively faster and more effective. 2) Scaling K-means to massive data is relatively easy with respect to the algorithm’s simplicity and iterative nature \cite{bahmani2012scalable}.

To perform K-means, we need the following parameters: 1) The distance function used to compute the distance between two points and the means of cluster centers: In this study, we use the distance function introduced in previous section as the distance function. 2) The selection of the number of clusters: In our experiment, we explored how to adequately set K for our dataset to gain the best performance. Since there is no perfect mathematical criterion exists \cite{jain2010data}, we experimented on how to set a best range of K values that could lead to better performance in the K-means approach. In order to implement the clustering method, we adopted the K-means function in Natural Language Toolkit \cite{loper2002nltk}, a leading platform for building Python programs to work with human language data. The K-means toolkit provided the flexibility for programmers to use their own distance function instead of Euclidean distance.

\subsubsection{STREAMCUBE Clustering}

STREAMCUBE used a single-pass hashtag clustering algorithm, shown in figure \ref{fig:algorithm_hashtag_cluster_static}. For each new hashtag, the algorithm first used a nearest-neighbor (shown in figure \ref{fig:algorithm_nearest_neighbor}) function to find the existing cluster nearest to the hashtag. The algorithm then checked the absorbing condition to decide if the hashtag should be absorbed into the nearest cluster. If the distance between the hashtag and the nearest cluster was greater than the cluster’s minimum threshold (i.e. the nearest distance between the cluster and any other clusters), the hashtag initialized a new cluster; Otherwise the hashtag was absorbed by the cluster.

\psfigx{algorithm_hashtag_cluster_static}{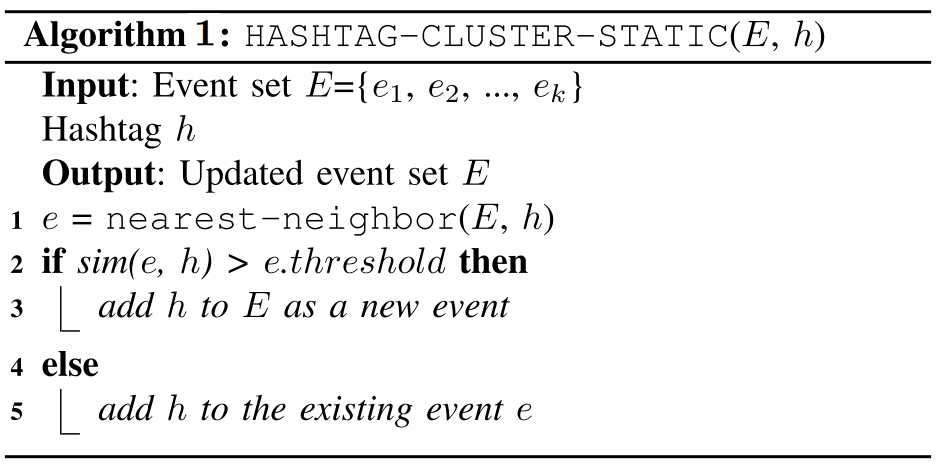}{0}{HASHTAG-CLUSTER-STATIC algorithm  \cite{feng2015streamcube}}{h}{3}

\psfigx{algorithm_nearest_neighbor}{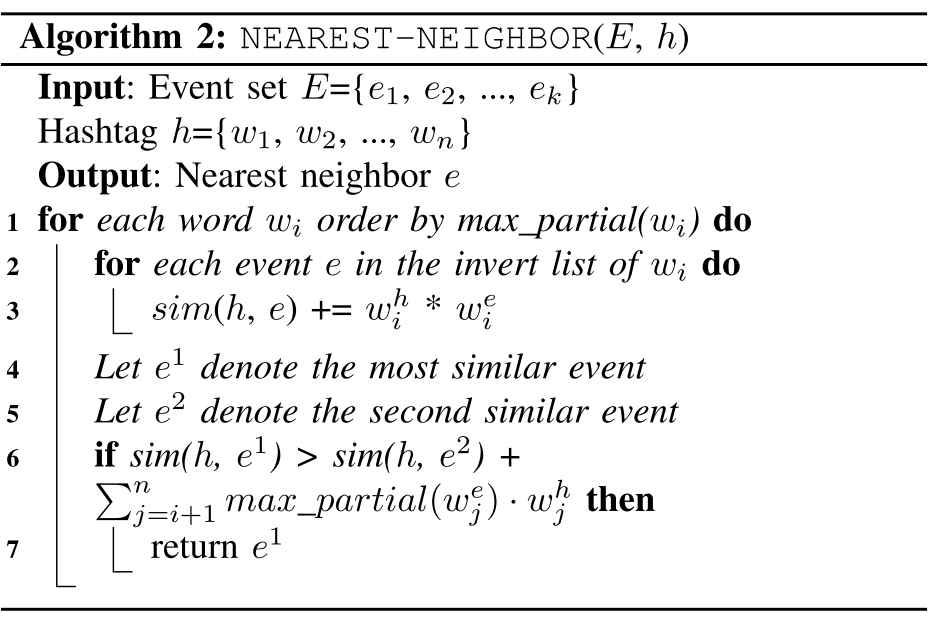}{0}{NEAREST-NEIGHBOR algorithm \cite{feng2015streamcube}}{h}{3}

\subsection{Data Analysis}

In terms of cluster analysis, there is no best measure for evaluating the cluster quality \cite{bhatnagar2010robust}. However, a mix of internal and external quality criteria provides us a comprehensive view to evaluate the clustering approaches. Therefore, we adopted two widely used metrics: Purity \cite{zhao2001criterion} as an external criterion and normalized mutual information (NMI) \cite{strehl2002cluster} as an internal criterion to evaluate the quality of the clustering results.

Purity is an external quality criterion and is used when classes in the data are known. It measures the extent that if the documents in a cluster are from primarily one specific class. Given there are $k$ clusters formed by total n documents that each document was labeled by one of $I$ classes. The Purity of $r$-th cluster $S_{r}$ with size $n_{r}$ is defined as the equation

$$ P(S_{r}) = \frac{1}{n_{r}}max_{i}(n_{r}^{i}), \forall i \in {1, 2, ..., |I|} $$ where $n_{r}^{i}$ is the number of documents of the $i$-th class that were assigned to the $r$-th cluster. The overall Purity of the clustering solution is obtained as a weighted sum of the individual cluster purities and is given by the equation

$$ Purity = \sum_{r=1}^{k}\frac{n_{r}}{n}P(S_{r}) $$ In general, the larger the values of Purity, the better the clustering solution is considered to be.

NMI is an internal quality criterion and captures the commonality between two clustering approaches. It provides an indication of the shared information between a pair of clusters. Given X and Y be the random variables described by the cluster labeling $\lambda^{(a)}$ and $\lambda^{(b)}$, with $k^{(a)}$ and $k^{(b)}$ respectively. Let $I(X,Y)$ denote the mutual information between $X$ and $Y$, $H(X)$, $H(Y)$ denote the entropy of $X$, $Y$. The equation of NMI is as the equation

$$ NMI(X,Y) = \frac{I(X,Y)}{\sqrt{H(X)H(Y)}} $$ The value of NMI is a fraction between 0 and 1, with 0 indicating that the two clusters do not shared the same information and 1 indicating that the two clusters are exactly the same.
\section{Experiments}

\subsection{Data Collection}

We collected 11,884,448 tweets from November 13, 2015 to November 17, 2015 for the Paris Attacks. The keyword list for collecting the tweets contained the following keywords: 'paris', 'attack', 'Gunmen', 'Bataclan', 'gunfire', 'hostage', 'Les Halles','Belle Equipe','Petite Cambodge', 'le Carillon’. We then filtered those tweets without text content, created time or geolocation to ensure the collected tweets did not lack any information we needed. Geolocation parameter is beyond the scope of this thesis. There were 20,514 tweets with 8,616 different hashtags in our tweet collection after filtering those without geolocation. In the following sections, we describe the design of our experiments from three different perspectives: 1) Comparing K-means to STREAMCUBE, with STREAMCUBE as the ground truth, 2) comparing K-means to STREAMCUBE, with human serving as the ground truth, and 3) finding better K values for K-means, with human serving as the ground truth.

\subsection{Compare K-means to STREAMCUBE (STREAMCUBE as the ground truth)}

In the first experiment, we compared the differences of clustering results between the K-means approach and the original STREAMCUBE, with STREAMCUBE as the ground truth. We did not consider which approach was better but investigated the commonality between the two approaches. First, we followed the group setting of STREAMCUBE to group the collected tweets by their created time into 6-hours, 12-hours, and 24-hours groups respectively. The original reason for this setting is because STREAMCUBE only keeps events from the last six hours in memory for increment updates in their online system. The historical data are fixed and flushed into disk-based storage \cite{feng2015streamcube}. Once every six-hours data go into disk-based storage, the system merges two six-hours data as a 12-hours data. The merge rule applied for the rest of the levels. Since their coarsest granularity is a day, the merge rule stops for 24-hours data. Although we did not aim to build a realtime system, we followed their setting to ensure the performance of STREAMCUBE was not influenced by a different group setting from its original. Second, we used STREAMCUBE to cluster the tweet groups. Since STREAMCUBE generated a dynamic number of clusters for every tweet group, we recorded the numbers of clusters for all tweet groups in order to use the numbers as the K values in the K-means approach. Third, we performed the K-means approach to cluster the tweet groups. Finally, we took the clustering results of STREAMCUBE as the ground truth and the clustering results of the K-means approach as the predictions to calculate the NMI and Purity scores. 

Tables \ref{tab:nmi_purity_6_hrs}, \ref{tab:nmi_purity_12_hrs} and \ref{tab:nmi_purity_24_hrs} listed the Purity and NMI scores of every 6, 12, 24 hours respectively. In table \ref{tab:nmi_purity_6_hrs}, the Purity scores showed that over 70\% of clusters generated by the K-means approach can be matched to corresponding clusters generated by STREAMCUBE, and the NMI scores showed the commonality between the results of the two clustering approaches are 57.8\% for 24-hour groups, 69.6\% for 12-hours groups, and 69.9\% for 6-hours groups respectively. The K-means approach and STREAMCUBE did share a large portion of similar clustering results, but some significant performance differences are worth investigating. To further understand the differences, we designed the experiment in section 4.3 to use human-labeled tweets for comparing the two approaches.

\begin{table}
\caption{The NMI and Purity scores of every 6 hours.}
\label{tab:nmi_purity_6_hrs}
\begin{center}
\begin{tabular}{ c c c c c c }
\hline
\bf Date & \bf Hour range & \bf Number of tweets & \bf Number of clusters & \bf Purity & \bf NMI \\
\hline
2015/11/13 & 18:00 - 24:00 & 783 & 8 & 83.3\% & 74.8\% \\
2015/11/14 & 0:00 - 6:00 & 1214 & 9 & 87.8\% & 78.8\% \\
2015/11/14 & 6:00 - 12:00 & 1038 & 6 & 71.4\% & 51.3\% \\
2015/11/14 & 12:00 - 18:00 & 1274 & 29 & 66.1\% & 79.0\% \\
2015/11/14 & 18:00 - 24:00 & 1848 & 11 & 75.3\% & 59.7\% \\
2015/11/15 & 0:00 - 6:00 & 1262 & 5 & 73.8\% & 63.6\% \\
2015/11/15 & 6:00 - 12:00 & 1451 & 13 & 77.5\% & 69.1\% \\
2015/11/15 & 12:00 - 18:00 & 1645 & 25 & 68.4\% & 74.3\% \\
2015/11/15 & 18:00 - 24:00 & 1302 & 6 & 73.2\% & 60.1\% \\
2015/11/16 & 0:00 - 6:00 & 1275 & 4 & 75.5\% & 63.7\% \\
2015/11/16 & 6:00 - 12:00 & 1598 & 25 & 67.2\% & 74.3\% \\
2015/11/16 & 12:00 - 18:00 & 1718 & 41 & 68.8\% & 80.3\% \\
2015/11/16 & 18:00 - 24:00 & 808 & 11 & 84.1\% & 76.1\% \\
2015/11/17 & 0:00 - 6:00 & 855 & 5 & 87.5\% & 83.2\% \\
2015/11/17 & 6:00 - 12:00 & 431 & 5 & 88.9\% & 79.9\% \\
2015/11/17 & 12:00 - 18:00 & 1191 & 6 & 63.8\% & 69.6\% \\
2015/11/17 & 18:00 - 24:00 & 821 & 5 & 63.4\% & 51.1\% \\
\hline 
\end{tabular}
\end{center}
\end{table}
\begin{table}
\caption{The NMI and Purity scores of every 12 hours.}
\label{tab:nmi_purity_12_hrs}
\begin{center}
\begin{tabular}{ c c c c c c }
\hline
\bf Date & \bf Hour range & \bf Number of tweets & \bf Number of clusters & \bf Purity & \bf NMI \\
\hline
2015/11/13 & 12:00 - 24:00 & 783 & 8 & 83.3\% & 74.8\% \\
2015/11/14 & 0:00 - 12:00 & 2252 & 27 & 75.0\% & 76.8\% \\
2015/11/14 & 12:00 - 24:00 & 3122 & 17 & 72.0\% & 61.8\% \\
2015/11/15 & 0:00 - 12:00 & 2713 & 11 & 64.1\% & 49.5\% \\
2015/11/15 & 12:00 - 24:00 & 2947 & 22 & 70.9\% & 65.6\% \\
2015/11/16 & 0:00 - 12:00 & 2873 & 40 & 72.6\% & 79.4\% \\
2015/11/16 & 12:00 - 24:00 & 2526 & 41 & 65.4\% & 73.0\% \\
2015/11/17 & 0:00 - 12:00 & 1286 & 8 & 80.9\% & 68.5\% \\
2015/11/17 & 12:00 - 24:00 & 2012 & 22 & 72.1\% & 77.0\% \\
\hline 
\end{tabular}
\end{center}
\end{table}
\begin{table}
\caption{The NMI and Purity scores of every 24 hours.}
\label{tab:nmi_purity_24_hrs}
\begin{center}
\begin{tabular}{ c c c c c c }
\hline
\bf Date & \bf Hour range & \bf Number of tweets & \bf Number of clusters & \bf Purity & \bf NMI \\
\hline
2015/11/13 & 783 & 8 & 83.3\% & 74.8\% \\
2015/11/14 & 5374 & 26 & 67.6\% & 56.2\% \\
2015/11/15 & 5660 & 2 & 99.3\% & 1.0\% \\
2015/11/16 & 5399 & 84 & 68.5\% & 78.5\% \\
2015/11/17 & 3298 & 46 & 67.1\% & 78.3\% \\
\hline 
\end{tabular}
\end{center}
\end{table}
\begin{table}
\caption{Summary of table \ref{tab:nmi_purity_6_hrs}, \ref{tab:nmi_purity_12_hrs} and \ref{tab:nmi_purity_24_hrs}}
\label{tab:nmi_purity_summary}
\begin{center}
\begin{tabular}{ c c c c }
\hline
\bf Hour range & \bf Avg. number of clusters & \bf Avg. Purity & \bf Avg. NMI \\
\hline
6 & 13.3 & 75.1\% & 69.9\% \\
12 & 21.8 & 72.9\% & 69.6\% \\
24 & 33.2 & 77.2\% & 57.8\% \\
\hline 
\end{tabular}
\end{center}
\end{table}

\subsection{Compare K-means to STREAMCUBE (human serving as the ground truth)}

We further compared the performance between the K-means approach and STREAMCUBE by using human-labeled tweets as the ground truth. First, we randomly selected 200 from 3298 tweets which contained 170 hashtags and 7,185 unigrams, collected on November 17, 2015. Second, we asked a human subject (a graduate student) to manually label categories for each of the 200 tweets. We instructed the subject to choose any text he wanted to label the tweets, but to use only one label for each tweet. The subject used six different labels in the labeling task: “Travel”, “Terrorism”, “Pray”, “Life”, “Hiring”, and “Others”. The label distribution of the 200 tweets is shown in figure \ref{fig:distribution_human_labeled_tweets}. Third, we used similar steps in section 4.2 to performed clustering on the tweets collected on November 17, 2015. We generated the clusters of STREAMCUBE and the clusters of the K-means approach respectively. Fourth, for each of the two cluster sets, we extracted the 200 labeled tweets and kept the cluster information from them. Thus, we had the clustering results of the 200 labeled tweets generated by the two approaches respectively, and we had the human-labeled information of the 200 tweets as the ground truth. Finally, we calculated the NMI and Purity scores for the K-means approach and STREAMCUBE respectively. Tables \ref{tab:hashtags_top_10_kmeans} and \ref{tab:hashtags_top_10_streamcube} list the hashtags of the top 10 large clusters generated by the K-means approach and by STREAMCUBE. Table \ref{tab:performance_compare} was the performance comparison between the K-means approach and STREAMCUBE. We have shown that the K-means approach performed better than STREAMCUBE on both the Purity and NMI scores given the same number of clusters. 

\psfigx{distribution_human_labeled_tweets}{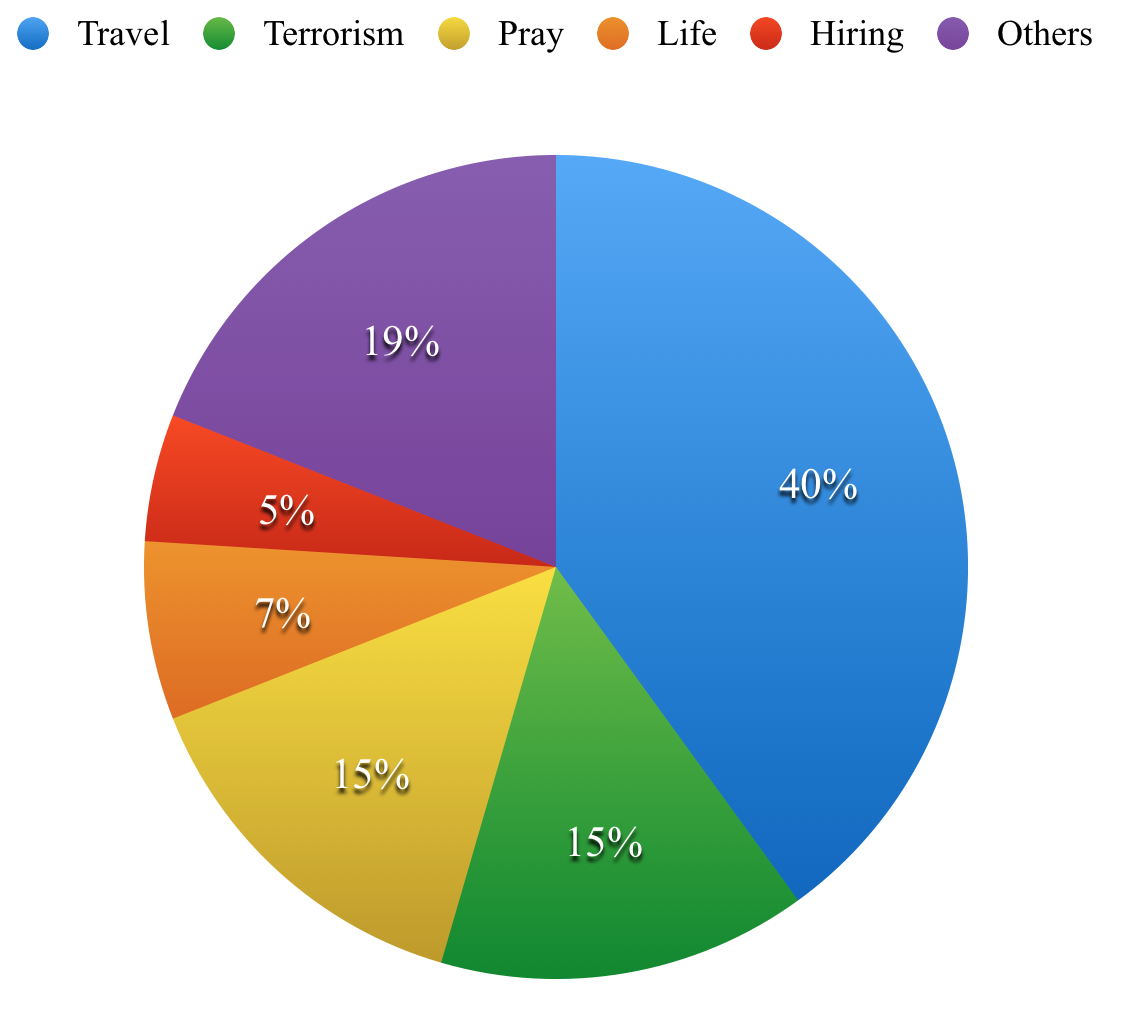}{0}{The distribution of the human labeled tweets}{h}{2}

\begin{table}
\caption{Hashtags of the top 10 large clusters generated by the K-means approach}
\label{tab:hashtags_top_10_kmeans}
\begin{center}
\begin{tabularx}{\textwidth}{ c X }
\hline
\multicolumn{2}{c}{\textbf{K-means}}\\
\hline
1 & parisattacks, igersparis, french, love, france, toulouse, city, picoftheday, photooftheday, pray, disneyland, fluctuatnecmergitur, europe, jesuisparis, peace, charliehebdo, parisian, tbt, view, prayforparis \\
2 & SONIC, CareerArc, Retail, Lebanon, Job, job, Veterans, ExpediaJobs, LEBANON, CustomerService, Jobs, Sales, Hiring, Hospitality \\
3 & Stigmabase, peaceforparis, informatique, vscocam, tb, movie, hope, instagood, stage, vsco, friends \\
4 & bomb, ParisAttacks, parismaville, Adidas, portrait, COP21, jesuisenterrasse, quiz, Montemartre \\
5 & TourEiffel, WeLoveParis, EiffelTower, MisterJoeCity, ILoveParis, Montmartre, France, Paris18, DirectLive \\
6 & 2DaysTilIKWYDLS, StreamMadeInTheAM, PrayForSyria, MTVStars, playpurpose, adtechNZ, SiyaKeRam, maritime \\
7 & blue, frenchlife, me, parisstreet, iloveparis, ootd, metro, parisjetaime \\
8 & tousaubistrot, concorde, attentat, parisattack, hommage, republique, placedelarepublique \\
9 & selfies, streetlife, selfiewithart, streetart, photography, parisnights \\
10 & london, football, huaweishot, Wembley, huawei, wembley \\
\hline 
\end{tabularx}
\end{center}
\end{table}

\begin{table}
\caption{Hashtags of the top 10 large clusters generated by STREAMCUBE}
\label{tab:hashtags_top_10_streamcube}
\begin{center}
\begin{tabularx}{\textwidth}{ c X }
\hline
\multicolumn{2}{c}{\textbf{STREAMCUBE}}\\
\hline
1 & JeSuisParis, travel, tousaubistrot, bomb, news, parisattacks, Stigmabase, vegas, pray, fluctuatnecmergitur, eiffeltower, jesuisparis, peace, prayforparis, PrayForParis \\
2 & PARIS, SONIC, CareerArc, Retail, Lebanon, hiring, Job, job, IT, Veterans, LEBANON, Transportation, Jobs, Hospitality \\
3 & selfies, frenchie, selfie, parisstreet, streetlife, iloveparis, selfiewithart, frenchart, streetart, photography, parisnights, parisjetaime \\
4 & 2DaysTilIKWYDLS, trndnl, StreamMadeInTheAM, PrayForSyria, MTVStars, playpurpose, adtechNZ, SiyaKeRam, maritime \\
5 & foodporn, ISIS, London, Adidas, Syria, COP21, Bataclan, ParisAttacks, movie \\
6 & expo, chezmatante, basket, creditmunicipal, inParis, art, villelumiere \\
7 & london, football, hnytwtr, huaweishot, Wembley, huawei, wembley \\
8 & liberté, blue, music, liberteegalitefraternite, shym, bercy, ootd \\
9 & attentat, parisattack, hommage, republique, freedom, placedelarepublique \\
10 & TourEiffel, WeLoveParis, EiffelTower, ILoveParis, France, DirectLive \\
\hline 
\end{tabularx}
\end{center}
\end{table}
\begin{table}
\caption{The performance comparison based on human-labeled categories}
\label{tab:performance_compare}
\begin{center}
\begin{tabular}{ c c c c }
\hline
\bf Clustering approach & \bf Number of clusters & \bf Purity & \bf NMI \\
\hline
K-means & 46 & 70.5\% & 35.6\% \\
STREAMCUBE & 46 & 67.1\% & 27.8\% \\
\hline 
\end{tabular}
\end{center}
\end{table}

\subsection{Find Better K Values for K-means (humans serving as the ground truth)}

We wanted to find the best K values for the K-means approach. Although we have shown the K-means approach could outperform STREAMCUBE in the previous experiments when using the same number of clusters, the K values of the K-means approach were chosen based on the results of STREAMCUBE. In this experiment, we performed experiments on the K-means approach and compared the performance between different K-means. We again used the 200 manually labeled tweets created in section 4.3 as the ground truth and the clustering results as the prediction. In table \ref{tab:performance_different_ks}, we performed the experiments for different K values and found that both the Purity and NMI scores were higher when the number of clusters is larger.

Although the clustering method is better when the Purity is greater, high Purity is easy to achieve when the number of clusters is large. Thus, we should not use Purity to trade off the quality of the clustering against the number of clusters \cite{schutze2009introduction}. The NMI scores reach 36\% and become stable when the number of clusters is greater than 20. Moreover, in table \ref{tab:performance_compare}, the Purity and NMI of STREAMCUBE was 67.1\% and 27.8\% while the number of cluster was 46. Results of table \ref{tab:performance_different_ks} show that once the number of clusters is greater than 20, the K-means approach could perform better than STREAMCUBE on both the Purity and NMI scores. Thus, according to our experiments, the K value for the K-means approach could be set at least greater than one tenth of the number of hashtags to achieve the performance better than STREAMCUBE.

\begin{table}
\caption{The performance of the K-means approach with different Ks}
\label{tab:performance_different_ks}
\begin{center}
\begin{tabular}{ c c c }
\hline
\bf Number of clusters & \bf Purity & \bf NMI \\
\hline
2 & 49.4\% & 15.2\% \\
5 & 53.9\% & 23.8\% \\
10 & 59.7\% & 31.1\% \\
15 & 66.5\% & 36.6\% \\
20 & 67.8\% & 36.6\% \\
30 & 70.0\% & 37.0\% \\
50 & 70.0\% & 35.4\% \\
100 & 73.8\% & 35.6\% \\
150 & 75.6\% & 36.2\% \\
170 & 76.8\% & 37.0\% \\
\hline 
\end{tabular}
\end{center}
\end{table}
\section{Conclusions}

In this study, we proposed an event detection approach that utilizes hashtags in tweets. We adopted the feature extraction used in STREAMCUBE for K-means clustering. To the best of our knowledge, this is the first study to extend the framework of STREAMCUBE by adopting different clustering algorithm to enhance the original STREAMCUBE. We collected the tweets related to the Paris Attack during November 13 to November 17, 2015 as our datasets and performed the following experiments: first, we compared the commonality and difference between the K-means approach and STREAMCUBE in the perspectives of Purity and NMI on a full set of over 20,000 tweets. Second, we collected manual labels for 200 randomly sampled tweets from a human subject and demonstrated that the K-means approach outperformed STREAMCUBE on the clustering results. Third, we further discussed how to set the K value for the K-means approach to lead to a better clustering performance.   


\bibliographystyle{splncs}
\bibliography{paper}
\end{document}